\documentclass[twocolumn,twocolappendix]{aastex631}
\usepackage{CJK}
\shorttitle{Metallicities of 20 Million Giant Stars Based on Gaia XP spectra}
\shortauthors{Yang et al.}

\usepackage{subfigure}
\usepackage{amsmath}
\usepackage{threeparttable}
\usepackage{array}

\makeatletter

\newcommand{\Rmnum}[1]{\expandafter\@slowromancap\romannumeral #1@}
\makeatother

\begin{document}
\begin{CJK*}{UTF8}{gbsn}

\title{Metallicities of 20 Million Giant Stars Based on Gaia XP spectra}

\correspondingauthor{Haibo Yuan}
\email{yuanhb@bnu.edu.cn}

\author[0000-0002-9824-0461]{Lin Yang (杨琳)}
\affiliation{Department of Cyber Security, Beijing Electronic Science and Technology Institute, Beijing, 100070, China}
\affiliation{College of Artificial Intelligence, Beijing Normal University No.19, Xinjiekouwai St, Haidian District, Beijing, 100875, P.R.China}

\author[0000-0003-2471-2363]{Haibo Yuan (苑海波)}
\affiliation{Institute for Frontiers in Astronomy and Astrophysics, Beijing Normal University, Beijing, 102206, China}
\affiliation{School of Physics and Astronomy, Beijing Normal University, Beijing, 100875, People's Republic of China}

\author[0000-0002-1259-0517]{Bowen Huang (黄博闻)}
\affiliation{Institute for Frontiers in Astronomy and Astrophysics, Beijing Normal University, Beijing, 102206, China}
\affiliation{School of Physics and Astronomy, Beijing Normal University, Beijing, 100875, People's Republic of China}

\author[0000-0003-1863-1268]{Ruoyi Zhang (张若羿)}
\affiliation{Institute for Frontiers in Astronomy and Astrophysics, Beijing Normal University, Beijing, 102206, China}
\affiliation{School of Physics and Astronomy, Beijing Normal University, Beijing, 100875, People's Republic of China}

\author[0000-0003-4573-6233]{Timothy C. Beers}
\affiliation{Department of Physics and Astronomy, University of Notre Dame, Notre Dame, IN 46556, USA}
\affiliation{Joint Institute for Nuclear Astrophysics -- Center for the Evolution of the Elements (JINA-CEE), USA}

\author[0000-0001-8424-1079]{Kai Xiao (肖凯)}
\affiliation{School of Astronomy and Space Science, University of Chinese Academy of Sciences, Beijing 100049, People's Republic of China}

\author[0000-0003-3535-504X]{Shuai Xu (徐帅)}
\affiliation{Institute for Frontiers in Astronomy and Astrophysics, Beijing Normal University, Beijing, 102206, China}
\affiliation{School of Physics and Astronomy, Beijing Normal University, Beijing, 100875, People's Republic of China}

\author[0000-0003-3250-2876]{Yang Huang (黄样)}
\affiliation{School of Astronomy and Space Science, University of Chinese Academy of Sciences, Beijing 100049, People's Republic of China}
\affiliation{Institute for Frontiers in Astronomy and Astrophysics, Beijing Normal University, Beijing, 102206, China}

\author[0000-0002-5818-8769]{Maosheng Xiang (向茂盛)}
\affiliation{National Astronomical Observatories, Chinese Academy of Sciences, 20A Datun Road, Chaoyang District, Beijing, China}
\affiliation{Institute for Frontiers in Astronomy and Astrophysics, Beijing Normal University, Beijing, 102206, China}
\author[0000-0001-9293-131X]{Meng Zhang (张萌)}
\affiliation{National Astronomical Observatories, Chinese Academy of Sciences, 20A Datun Road, Chaoyang District, Beijing, China}
\author[0009-0005-7743-6229]{Jinming Zhang (张津铭)}
\affiliation{Institute for Frontiers in Astronomy and Astrophysics, Beijing Normal University, Beijing, 102206, China}
\affiliation{School of Physics and Astronomy, Beijing Normal University, Beijing, 100875, People's Republic of China}

\begin{abstract}
We design an uncertainty-aware cost-sensitive neural network (UA-CSNet) to estimate metallicities from dereddened and corrected Gaia BP/RP (XP) spectra for giant stars. This method accounts for both stochastic errors in the input spectra and the imbalanced density distribution in [Fe/H] values. With a specialized architecture and training strategy, the UA-CSNet improves the precision of the predicted metallicities, especially for very metal-poor (VMP; $\rm [Fe/H] \leq -2.0$) stars.
With the PASTEL catalog as the training sample, our model can estimate metallicities down to $\rm [Fe/H] \sim -4$. We compare our estimates with a number of external catalogs and conduct tests using star clusters, finding overall good agreement. 
We also confirm that our estimates for VMP stars  are unaffected by carbon enhancement. Applying the UA-CSNet, we obtain reliable and precise metallicity estimates for approximately 20 million giant stars, including 360,000 VMP stars and 50,000 extremely metal-poor (EMP; $\rm [Fe/H] \leq -3.0$) stars. The resulting catalog is publicly available at https://doi.org/10.12149/101604. This work highlights the potential of low-resolution spectra for metallicity estimation and provides a valuable dataset for studying the formation and chemo-dynamical evolution of our Galaxy. 
\end{abstract}


\section{Introduction} \label{introduction}

Metallicity, often parameterized as [Fe/H], is a fundamental stellar parameter that plays a key role in studying not only the formation and evolution of stars, but also galaxies like the Milky Way (MW). 
Already executed, ongoing, and planned large-scale spectroscopic surveys, such as the Sloan Digital Sky Survey and the Sloan Extension for Galactic Understanding and Evolution (\citealp[SDSS/SEGUE;][]{York2000,Yanny2009,Rockosi2022}), the Radial Velocity Experiment (\citealp[RAVE][]{Steinmetz2006}), the Large Sky Area Multi-Object Fiber Spectroscopic Telescope (\citealp[LAMOST;][]{Luo2015}), the Apache Point Observatory Galactic Evolution Experiment (\citealp[APOGEE;][]{Majewski2017}), the 4m Multi-Object Spectroscopic Telescope (\citealp[4MOST;][]{de2019}), the GALactic Archaeology with HERMES (\citealp[GALAH;][]{Buder2018}), and the Gaia DR3 General Stellar Parametriser-spectroscopy (\citealp[Gaia GSP-Spec;][]{Recio2023}) based on the Gaia Radial Velocity Spectrometer (\citealp[RVS;][]{Katz2004,Cropper2018}) spectra have (or will) provided extensive spectroscopic data that yield precise metallicity estimates for many millions of stars. 

In addition to spectroscopic data, the European Space Agency's Gaia mission (\citealp{Gaia2016}) provides astrometric data with unprecedented precision for billions of stars, offering a new perspective on the chemo-dynamical evolution of the Milky Way. In its third data release, Gaia DR3 (\citealp{Gaia2023}) delivered approximately 220 million low-resolution BP and RP spectra (hereafter referred to as Gaia XP spectra), with a resolution of $R \sim$ 20--70 and covering a wavelength range of 336--1020 nm. The potential of Gaia XP spectra for metallicity estimation has been demonstrated in a number of studies (e.g., \citealp{Liu2012,Rix2022,Andrae2023a,Andrae2023b,Zhang2023,Fallows2024,Huang2025,Laroche2024,Leung2024,Li2024,Xylakis2024,Yao2024,Ye2024}).

Recently, data-driven methods have emerged as an effective approach to infer stellar metallicity from low-resolution spectra, leveraging real data from well-characterized stars for comparison. By training a model that incorporates constraints from theoretical spectral models, \cite{Xiang2019} estimated 16 elements for 6 million stars using LAMOST DR5 low-resolution spectra ($R \sim 1800$), achieving a typical internal abundance precision of 0.03--0.3 dex. In another example, \cite{Angelo2024} used a subset of well-characterized common stars observed by GALAH to train a model based on the Cannon (\citealp{Ness2015}) framework, deriving stellar labels from Gaia DR 3 RVS spectra. However, Gaia XP spectra are not like traditional spectroscopic data. Their extremely low resolution and high flux calibration precision make Gaia XP spectra more suitable for treatment as high-quality photometric data.

Data-driven methods 
are also widely used for metallicity estimates from photometric data, such as the random forest algorithm (\citealp{Galarza2022,Andrae2023a}),
Bayesian method (\citealp{Anders2019,Andrae2023b}), kernel principal component analysis (\citealp{Huang2024b}), and artificial neural
networks (ANNs; \citealp{Thomas2019,Whitten2019,Whitten2021,Yang2022}). In particular, ANNs, which excel at extracting stellar parameters from photometric colors, have become widely used. Unlike spectroscopy, which relies on local features such as spectral lines to quantify metallicity, photometric estimation depends on global characteristics. This makes multilayer perceptron (MLP)-based networks more suitable than convolutional neural networks (CNNs) when applying ANN algorithms for photometric metallicity estimation. Using mixed-bandwidth photometric colors from the Javalambre Photometric Local Universe Survey (\citealp[J-PLUS;][]{Cenarro2019}), \cite{Whitten2019} proposed the Stellar Photometric Index Network Explorer (SPHINX), an MLP-based network designed to derive metallicities of stars in the $T_{\rm eff}$ range of 4500--6200 K from J-PLUS DR1, achieving a typical precision of 0.25 dex. Later, \cite{Whitten2021} applied SPHINX to the Southern Photometric Local Universe Survey (\citealp[S-PLUS;][]{Mendes2019}) Data Release 2 (\citealp{Almeida2022}), obtaining metallicity estimates for over 700,000 stars with a typical precision of $\sim 0.25$ dex. However, the metallicities predicted by such ANN models can be uncertain due to input noise and incorrect model inference (\citealp{Psaros2023}). To address this, \cite{Fallows2022} proposed a neural network approach that employs dropout as a regularization technique to estimate metallicities and their uncertainties from eight photometric colors, sourced from the Gaia EDR3, 2MASS, and WISE surveys for red giant stars. They achieved a typical precision of $\rm \delta[Fe/H]\sim$0.15 dex for stars of $\rm -0.5<[Fe/H]<0.5$.

In a previous study (\citealp{Yang2022}), we demonstrated that the cost-sensitive neural network (CSNet) performs very well in delivering metallicities from high-quality photometric data, even for very metal-poor (VMP; $\rm [Fe/H] \leq -2.0$) stars, by giving extra attention to stars in poorly populated regions. Taking into account the varying S/Ns across different Gaia XP spectra, we developed CSNet as an uncertainty-aware model that on one hand incorporates errors in the Gaia XP spectra and on the other hand includes extremely metal-poor (EMP; $\rm [Fe/H] \leq -3.0$) stars during the training of the stellar parameter estimation model. This approach improves the overall precision and range of metallicity estimates down to [Fe/H] $\sim -4$ by fully utilizing prior observational errors and applying greater error penalties to rare star samples.

The paper is organized as follows: In Sections\,\ref{data} and \ref{method}, we describe the data and proposed method in this work. Section\,\ref{Result} presents the results of our model, along with a number of validations. The resulting metallicity catalog of 20 million giant stars is also included in Section\,\ref{Result}. We
summarize in Section\,\ref{summary}.

\section{Data} \label{data}

The datasets used in this work consist of Gaia DR3 XP spectra \citep{De2023}, 
as well as the PASTEL \citep{Soubiran2010,Soubiran2016} and Stellar Abundances for the Galactic Archeology \citep[SAGA;][]{Suda2008,Suda2011,Suda2017,Yamada2013} catalogs. The Gaia DR3 dataset provides stellar SEDs, while the PASTEL and SAGA catalogs offer the metallicity labels derived from high-resolution spectroscopy. 

\subsection{Corrected Gaia XP Spectra} 
\label{Processed Spectra}

Gaia is a European Space Agency mission, aiming to explore the kinematical, dynamical, and chemical structure and evolution of the MW through precise astrometry, photometry, and spectroscopy \citep{Gaia2016}. Its third data release, Gaia DR3 has provided approximately 220 million low-resolution XP spectra (Blue and Red Photometers, BP and RP), covering the optical to near-infrared wavelength range of 330 to 1050 nm \citep{Carrasco2021,De2023,Gaia2023}.
Gaia XP spectra are represented as a linear combination of Hermite functions, utilizing two sets of 55 basis functions for the BP and RP spectra, respectively. We used the \texttt{GaiaXPy} package \citep{Ruz2022} to transform the above Gaia BP and RP coefficients into default-sampled spectra and corresponding errors, covering the wavelength range of 330 to 1020 nm with a 2 nm sampling interval. 

Gaia XP spectra suffer systematic errors that depend on the normalized spectral energy distribution and $G$ magnitude, particularly in the blue where the sensitivity to metallicity is strongest, which in turn affect the accuracy and reliability of metallicity measurements. To address these errors, we applied a comprehensive correction for systematic wavelength errors to the Gaia XP spectra, within the ranges of approximately $-0.5<BP-RP<2.0$, $3.0<G<17.5$, and $E(B-V)<0.8$, using the Python package from \cite{Huang2024}. Accurate extinction correction is also crucial for obtaining precise metallicity measurements with the proposed method. Therefore, extinction corrections were carried out using the dust-reddening map from \cite{Schlegel1998} and the empirical extinction curve from \cite{Zhang2024}. 
Note that the extinction curve of \cite{Zhang2024} was constructed using the corrected 
$Gaia$ XP spectra of approximately 370,000 common sources from $Gaia$ DR3 and LAMOST DR7. \cite{Zhang2024} used 440 nm and 550 nm as approximations for the Johnson B and V bands, respectively, and the corresponding relationship is given by
\begin{equation}\label{Z}
    E(44-55)=0.961*E(B-V)_{SFD}+1.1e^{-3}
\end{equation}

To reduce spurious features in the non-truncated spectra, the corrected and dereddened Gaia XP spectra were re-sampled by calculating median flux values within specified wavelength ranges, resulting in spectra represented by 60 pixels. Then, the re-sampled spectra were scaled by dividing the mean flux value at the wavelength of 5500 \AA, which is free of strong stellar absorption features. The errors of the re-sampled spectra were computed as the root mean square (RMS) of the original errors within the same wavelength ranges used for the flux re-sampling. The Gaia XP spectra mentioned hereafter in this work refer to those corrected, dereddened, and then re-sampled ones.

\vskip 2cm
\subsection{PASTEL and SAGA Stellar Metallicities} 

The PASTEL catalog\footnote{\url{https://vizier.cds.unistra.fr/viz-bin/VizieR?-source=B/pastel}} compiles stellar atmospheric parameters derived by applying model atmospheres to high-resolution ($R \geq 25,000$) and high signal-to-noise ($S/N \geq 50$) spectra \citep{Soubiran2010,Soubiran2016}. To ensure the reliability of the determinations, only records from references published after 1986 were used in this work. The SAGA catalog\footnote{\url{http://saga.sci.hokudai.ac.jp/}} compiles stellar parameters of metal-poor stars from references published since 1994 \citep{Suda2008,Suda2011,Suda2017,Yamada2013}. After excluding stars in common with the PASTEL catalog, the metallicity values from the SAGA catalog were used to evaluate the performance of our trained model.

\subsection{Experimental Data Construction} 
\label{training and tesing sets}

\begin{figure*}
  \centering
  \includegraphics[width=0.8\textwidth]{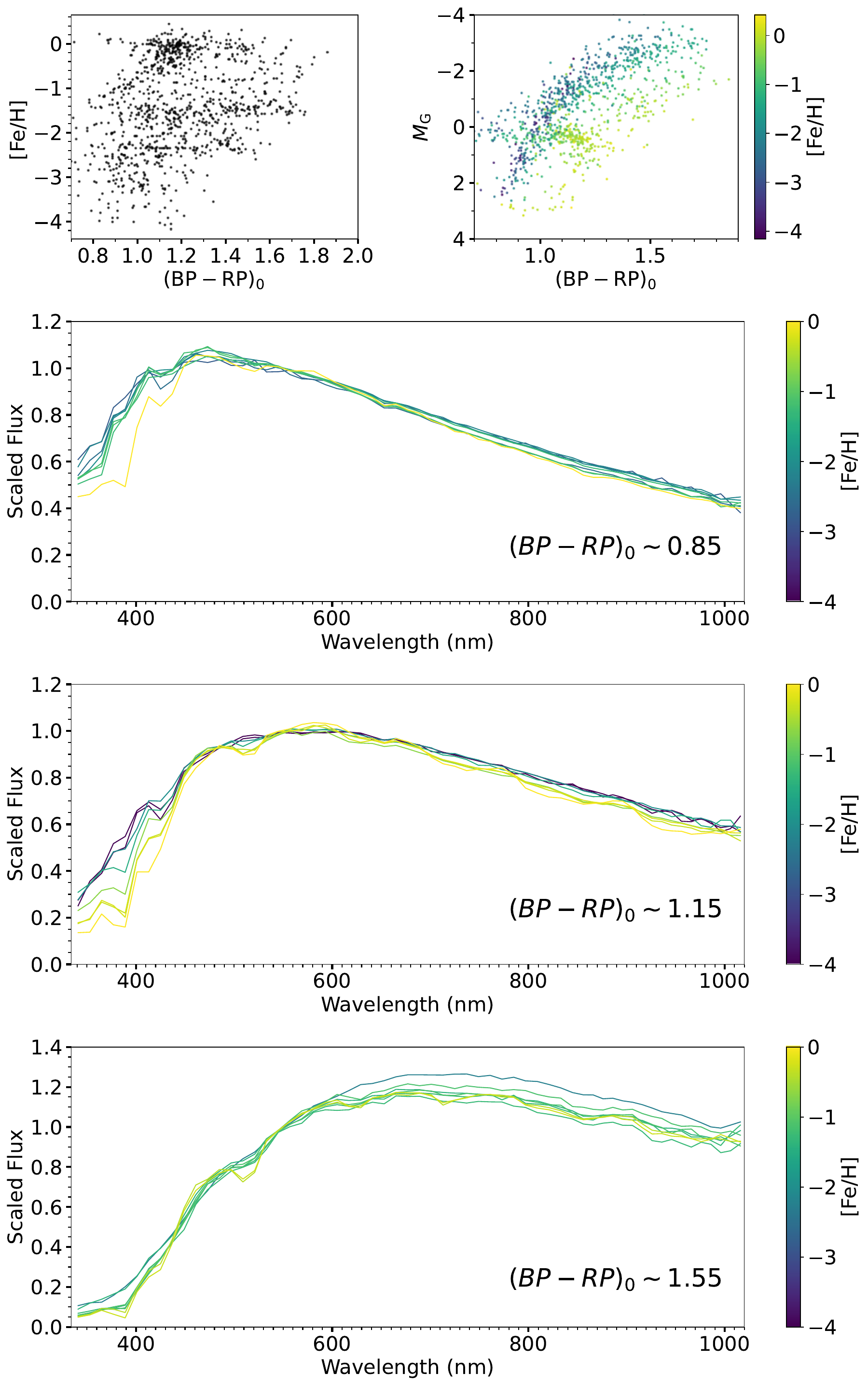}
  \caption{The top left and right panels show the distribution of the training set in the plane of 
  $(BP-RP)_0$ -- [Fe/H] and $(BP-RP)_0$ -- $M_G$, respectively. The bottom three panels each compare eight representative Gaia XP spectra from the training set that span a range of [Fe/H] values, but share similar $(BP-RP)_0$ colors of approximately 0.85, 1.15, and 1.55, respectively. The dots in the top right panel and the lines in the bottom three panels are color-coded by [Fe/H].}
  \label{fig:sample_space}
\end{figure*}

In this work, we focus on giant stars.  
By cross-matching Gaia XP spectra with the PASTEL and SAGA catalogs, we obtained 13,638 and 1,195 stars in common, respectively.  
The following cuts were applied to ensure the use of high-quality data: 
\begin{itemize}
   \item [1.] \texttt{phot}$\_$\texttt{bp}$\_$\texttt{rp}$\_$\texttt{excess}$\_$\texttt{factor} $ < 1.15+0.19*(BP-RP)_0^2$ to exclude stars of inaccurate {\it Gaia} data (\citealp{Xu2022});
    \item [2.]
    $0.7<(BP-RP)_0<1.9$ and $M_G<-(BP-RP)_0^2+6.5*(BP-RP)_0-1.8$ to select red giant stars (\citealp{Xu2022}), where $(BP-RP)_0$ is the dereddened $BP-RP$ color; 
    \item [3.]
    The distance to the Galactic plane $|Z|$ larger than 0.2 kpc;
    \item [4.]
    $|b|>20^{\circ}$ to avoid high extinction in the low Galactic latitude region; 
    \item [5.]
    $E(B-V)_{\rm SFD}$ $<$ 0.2 mag to exclude stars of high extinction, where $E(B-V)_{\rm SFD}$ is from the SFD98 map \citep{Schlegel1998};
    \item [6.]
    Dereddened $G$-magnitude $G_0>6$.
\end{itemize}

Finally, a training set of 1012 sample stars from PASTEL and a testing set of 649 sample stars from SAGA were selected. Note that there are no common sources between the training and testing sets. The top left and right panels of Figure \ref{fig:sample_space} show the distribution of the training set in the $(BP-RP)_0$ -- [Fe/H] plane and color-magnitude diagram, respectively. 
The bottom three panels each compare eight representative Gaia XP spectra from the training set that span a range of [Fe/H] values, but share similar $(BP-RP)_0$ colors of approximately 0.85, 1.15, and 1.55, respectively. It can be observed that the variation range depends on the $BP-RP$ color. For redder stars, variations in the blue wavelength range are more pronounced, indicating a higher sensitivity of the Gaia XP spectra to metallicity.

\section{Method} \label{method}

\begin{figure*}
  \centering
\includegraphics[width=0.8\textwidth]{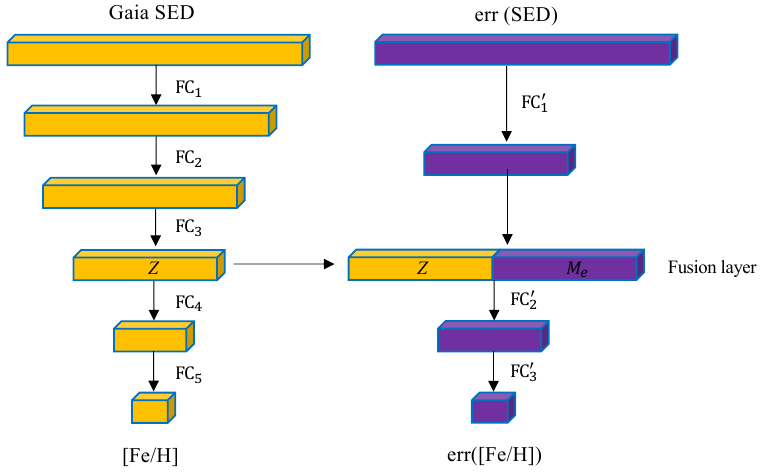}
  \caption{The structure of the proposed uncertainty-aware CSNet.}
  \label{fig:network}
\end{figure*}

In this work, we focus on metal-poor stars, which account for a relatively small proportion of the  training sample. To mitigate the impact caused by this data imbalance, we followed the cost-sensitive neural network approach (\citealp[CSnet;][]{Yang2022}), assigning higher weights to less populated regions in the [Fe/H] space. Considering the input spectra across varying levels of the noise, we developed the uncertainty aware CSNet (UA-CSNet), which estimates metallicities and their corresponding errors by providing a probability distribution with the mean and variance of a Gaussian. The variance consists of both epistemic and aleatory uncertainties. Epistemic uncertainty arises from a lack of knowledge about the model's structure and parameters, often due to limited training samples in specific regions. Aleatory uncertainty, on the other hand, results from inherent variability in the input data. Aleatory uncertainty can be further classified into two types: homoscedastic and heteroscedastic. Given the varying observation errors in Gaia XP spectra and the imbalanced distribution of [Fe/H] values in the training set, we model heteroscedastic aleatory uncertainty and epistemic uncertainty in this work. It is important to note that modeling the variance in predicted [Fe/H] is an unsupervised learning task. This approach not only quantifies prediction errors within the $G$ range of the training data but also provides an implicit and dynamic weight for input data during the training process. The proposed model converged during the training process using a customized loss function.

\subsection{The Uncertainty-aware CSNet} \label{CSNet}

As shown in Figure \ref{fig:network}, the uncertainty-aware CSNet is divided into two branches: the [Fe/H] branch and the uncertainty branch. The [Fe/H] branch takes Gaia XP spectra as the input vector $X$ and returns a latent representation $Z$ along with the predicted means $\mu$ of the [Fe/H] values. The uncertainty branch takes the concatenation of the corresponding spectrum errors $X_{err}$ and $Z$ as the input vector and returns the predicted heteroscedastic uncertainties $\sigma$ of the values of [Fe/H]. 

For the [Fe/H] branch, the expanded MLP network performs two functions: (1) extracting a deep feature map $Z\in \mathbb{R}^{C}$ that approximately contains all information of stellar atmospheric parameters and
elemental abundances, and (2) mapping this feature map to the [Fe/H] values. The latent representation $Z$ is calculated by a linear combination of $m$ non-linear functions:
\begin{equation}\label{Z}
    Z=g([\phi_1(X),\phi_2(X),...,\phi_m(X)]w^T)
\end{equation}
where $g(\bullet)$ denotes the rectified linear unit activation function (ReLU), and $w$ is a $m$ dimensional vector of learnable coefficients. An appropriate value for 
$m$ is slightly greater than the number of stellar parameters represented in the Gaia XP spectra.
Next, a sequence of fully connected layers (FCs) maps $Z$ to [Fe/H] by applying the following nonlinear functions:
\begin{equation}\label{Y}
Y=\Phi_{FC}(Z)
\end{equation}
where $\Phi(\bullet)$ is a linear combination of nonlinear functions from fully connected layers.

For the uncertainty branch, the input feature map consists of two parts: (1) the feature map $M_{e}\in \mathbb{R}^{C}$, generated by the full connection layers, which encodes information about heteroscedastic aleatory uncertainty, and (2) the feature map \( Z \in \mathbb{R}^C \) from the [Fe/H] branch, which captures information about epistemic uncertainty. These two types of uncertainty are then merged as follows:
\begin{equation}\label{Me}
    M_e^{'}=[Z;M_e]
\end{equation}
Then, the uncertainty of the [Fe/H] value is determined as 
\begin{equation}\label{sigma}
\delta Y_{\rm model}=\Phi_{e}(M_e)
\end{equation}

\subsection{The Model Objective}

We consider the mapping from the input spectra $X$ to the [Fe/H] values $Y$ and their corresponding errors $Y_{err}$ as a parameter estimation problem for a Gaussian distribution. The predicted [Fe/H] value is defined as:
\begin{equation}
    Y_i \sim \mathcal N(\Phi_{FC} (Z),\delta Y_i ^2)
\end{equation}
To speed up the model convergence, lower boundaries $\epsilon$ are imposed on the $Y_{err}$ values as
\begin{equation}\label{epsilon}
    \delta Y_i = max (\epsilon , \delta Y_{\rm i,model}^2)
\end{equation}
The probability of $Y_i$ given a spectra input $F_i$ can be approximated by a Gaussian distribution
\begin{equation}
    p(Y_i|X_i) = \frac{1}{\sqrt{2\pi}\sigma_i Y_i}\times {\rm exp}\left(-\frac{(Y_i-\Phi(Z))^2}{2\sigma_i^{2}}\right)
\end{equation}
In maximum likelihood inference, we define the likelihood as
\begin{equation}\label{likelihood}
    p(Y_1,...,Y_k,|F)=p(Y_1|F)...p(Y_k|X)
\end{equation}
To train $\sigma$ and $Y$, we optimize the parameters through minimizing the following loss function $\mathcal L$, which is the negative logarithm of the Eq.\,(\ref{likelihood}) likelihood
\begin{equation}\label{loss}
    \mathcal {L}=\sum_{i=1}^{n}\left(\frac{(Y_i-\Phi_{FC} (Z))^2}{2\sigma_i^2}+\log \sigma_i\right)
\end{equation}

Cost-sensitive learning and $L_2$ regularization are incorporated into our model to address the issue of imbalanced data and mitigate overfitting. The input data set $X$ is divided into M*N bins in the $(BP-RP)_0$--[Fe/H] space. Next, different costs are assigned to samples in each bin of the histogram according to the following rule:
\begin{equation} \label{frequency distribution histogram}
  c(x_i)=(\frac{f_n(x_i)}{max(f_n(X))})^{-\gamma},
\end{equation}
where $x\in X$ is a sample, $f_n(x_i)$ is a function to calculate the frequency of a group including $x$ in the histogram, $\gamma>0$ controls the difference degree of the cost among different bins. Thus, the final cost function is 
\begin{equation} \label{cost loss}
  \mathcal {L}=c(x_i)\sum_{i=1}^{n}\left(\frac{(Y_i-\Phi_{FC} (Z))^2}{2\sigma_i^2}+\log \sigma_i\right)
\end{equation}

\section{Results} \label{Result}

The uncertainty-aware CSNet was trained using Tensorflow 2.14.0, CUDA 11.8, and cuDNN 8, with GPU acceleration. The experiments were conducted on a server equipped with two Intel Xeon Gold 6238 CPUs and two NVIDIA GeForce RTX 4090 GPUs. To assess the accuracy and reliability of the results from our model, we performed extensive experiments evaluating the estimated metallicities from various perspectives. After introducing the experimental setup, we trained and tested our model using metallicities from the PASTEL and SAGA catalogs and compared our results with those from previous Gaia catalogs and other surveys for stars in common. To further validate the reliability of our results, we tested on star clusters and ensured that carbon enhancement did not influence the results.

\subsection{Hyperparameter Setup} \label{Hyperparameter Setup}

Before training the uncertainty-aware CSNet, hyperparameters related to both the architecture and optimization were set manually. Architecture-related hyperparameters include the number of neurons and hidden layers of fully connected layers, the activation function, and the weight initialization method. For the [Fe/H] branch, the architecture-related hyperparameters are summarized in Table\,\ref{table:table1}. They were
set based on the hyperparameters used in cost-sensitive neural networks \citep[CSNet;][]{Yang2022} and the uncertainty-aware residual attention network \citep[UaRA-net;][]{Yang2024}. To train the errors of the predicted [Fe/H] values through the uncertainty branch, two hidden layers with 16 and 8 neurons, along with a concatenation layer, were used to extract deep representations corresponding to the spectral errors and stellar parameters.

The optimization-related hyperparameters include the batch size of the training samples $s$, the number of training iterations $epoch$, the learning rate $\eta$, the $(BP-RP)_0$--[Fe/H] interval (M, N), the predictive error floor $\epsilon$ in the cost function Eq.\,(\ref{epsilon}), and the exponent of the weight assignment function  Eq.\,(\ref{frequency distribution histogram}). Through experimentation, we found that the following default settings yielded good performance: $s=512$, $epoch=30000$, $\eta=0.001$, $M=10$, $N=15$, $\lambda=10^{-5}$, $\epsilon=0.02$, and $\gamma=0.5$.

\begin{table}[ht!]
\centering
\caption{Architecture-related Hyperparameters for Training [Fe/H] and Err([Fe/H])}
\label{table:table1}
\begin{tabular}{cccc}
\hline
\multicolumn{2}{c}{[Fe/H]}  & \multicolumn{2}{c}{err([Fe/H])} \\
\hline
 Layers &  Output size & Layers &  Output size  \\
\hline 
  Input &  60 & Input & 60 \\
  $\rm FC_1$ &  256 &--  &--\\
  $\rm FC_2$ &  64 &$\rm FC'_1$ &16\\
  $\rm FC_3$ &  16 & Concat &32\\
  $\rm FC_4$ &  8 & $\rm FC'_2$ & 8\\
  $\rm FC_5$ &  1 & $\rm FC'_3$ & 1\\    
\hline
\end{tabular}
\end{table}

\subsection{Performance} \label{Performance}
To avoid model extrapolation and ensure the reliability of the results, we restricted the application of the trained model to stars only within the same $\rm (BP-RP)_0$ range of 0.7 to 1.9 as the training set. 

\subsubsection{Metallicity Estimates on the Training and Testing Sets}
\begin{figure*}
  \centering
  \includegraphics[width=0.9\textwidth]{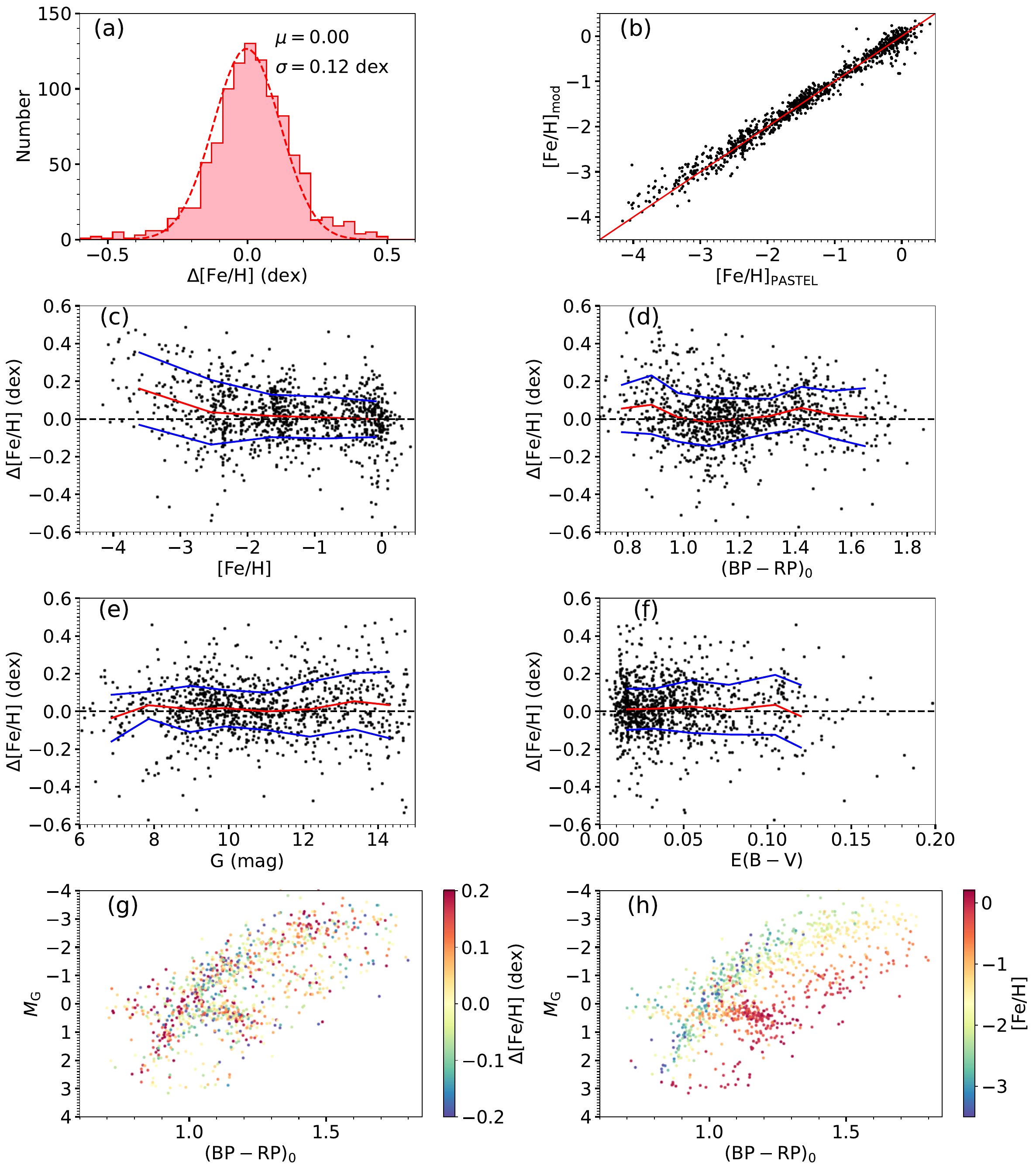}
  \caption{(a): Histogram of the residuals $\Delta$[Fe/H], with the mean and standard deviation values labeled. (b): Comparison of [Fe/H] values for common stars between our results and the PASTEL catalog. The red solid line is the one-to-one line. (c)--(f): Fitting residuals, as a function of [Fe/H], intrinsic $BP-RP$ color, $G$-magnitude, and $E(B-V)$. The red and blue solid lines indicate the mean deviation and 1$\sigma$ uncertainties, respectively. (g): The color-magnitude diagram, color-coded by $\Delta$[Fe/H]. (h) The distribution of the predicted [Fe/H] of the training set in the plane of $(BP-RP)_0$ -- $M_G$, color-coded by [Fe/H] values from our model.} 
  
  \label{fig:PASTEL}
\end{figure*}

\begin{figure*}
  \centering
  \includegraphics[width=0.9\textwidth]{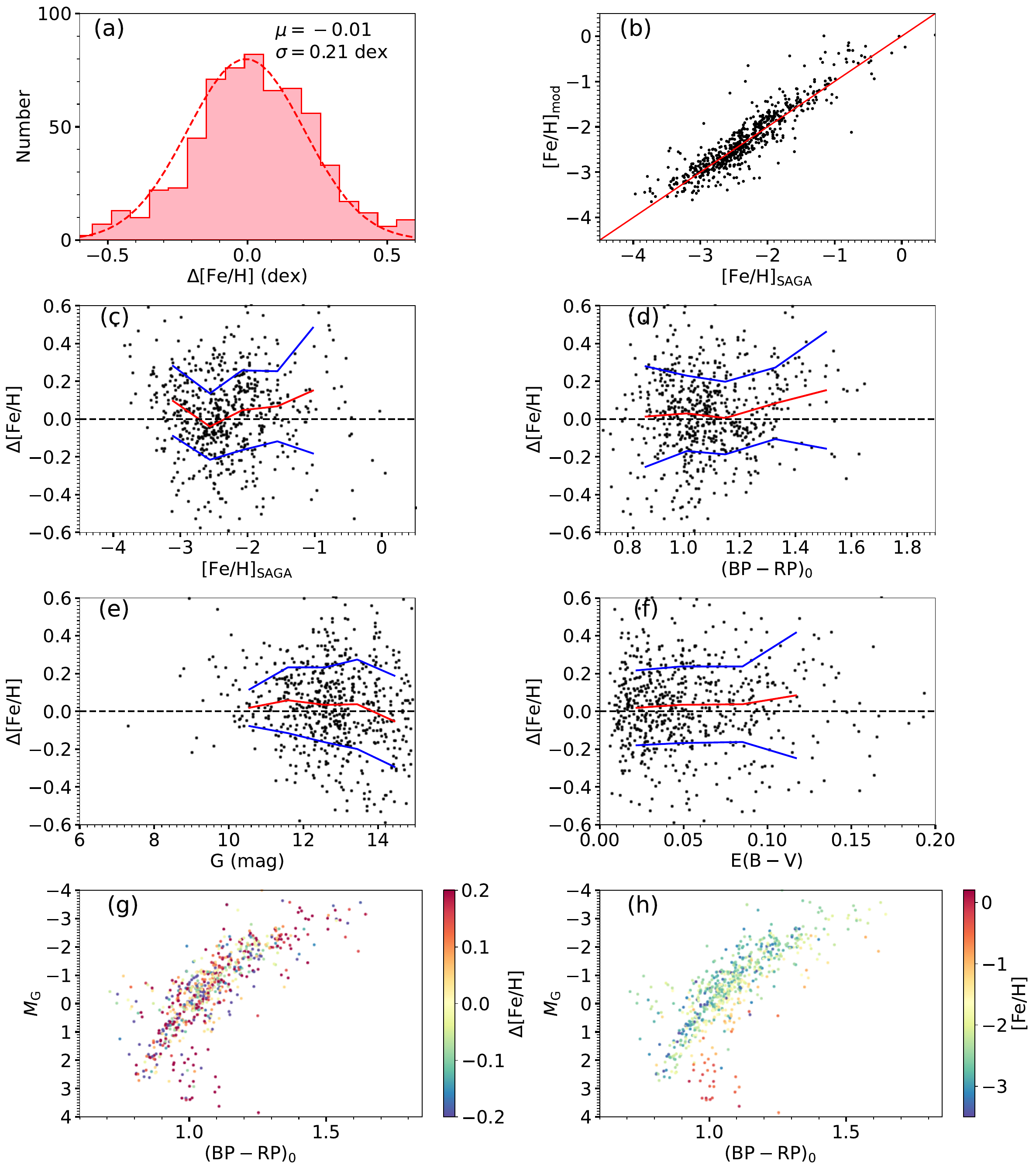}
  \caption{Similiar to Figure \ref{fig:PASTEL}, but for the SAGA catalog.}
  \label{fig:SAGA}
\end{figure*}

\begin{figure*}
  \centering
  \includegraphics[width=0.9\textwidth]{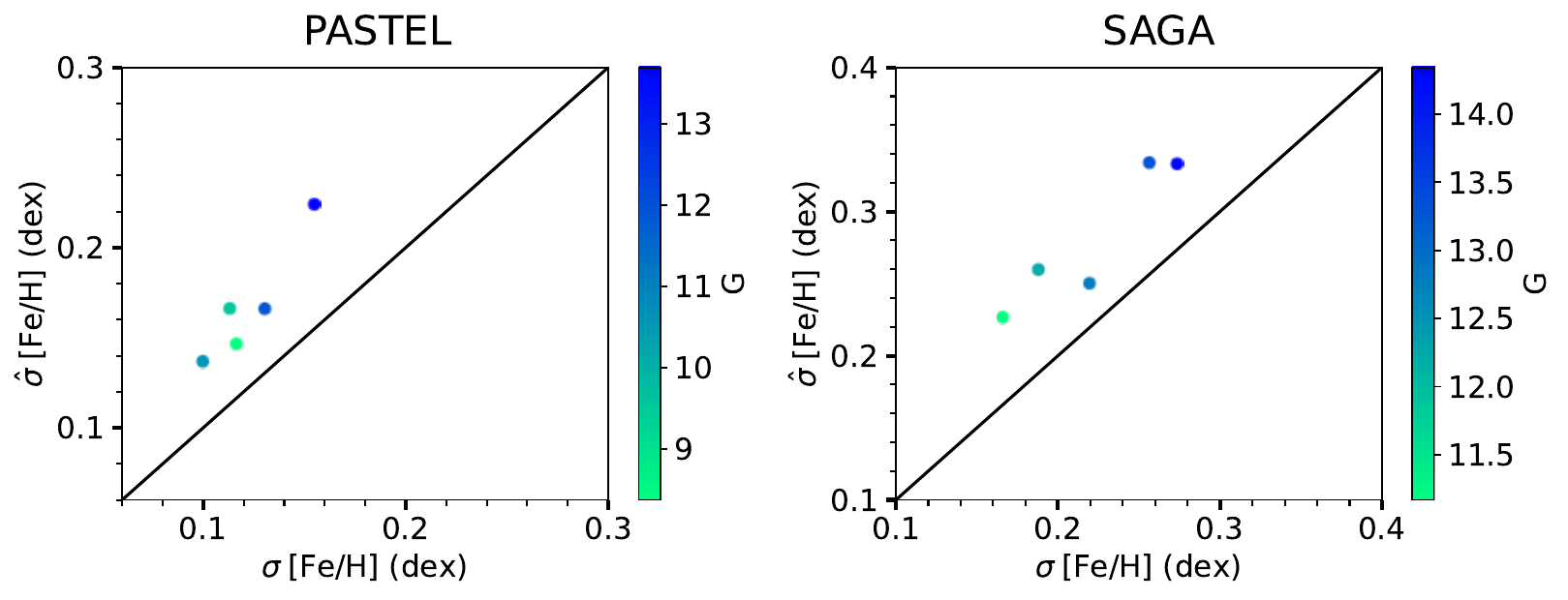}
  \caption{Comparisons between typical predicted errors (y-axis) from the uncertainty-aware CSNet and 1$\sigma$ uncertainties of the residuals (x-axis) for the PASTEL (left panel) and SAGA (right panel) catalogs in different $G$-magnitude bins.}
  \label{fig:err_branch}
\end{figure*}

To evaluate the accuracy of the metallicities derived from our model, we first compare our results with the PASTEL and SAGA catalogs, which are the training and testing sets, respectively. Note that the SAGA catalog was not involved in the model training, making this an independent validation. Figures \ref{fig:PASTEL} and \ref{fig:SAGA} plot the residuals for training and testing set, as a function of metallicity, intrinsic $BP-RP$ color, $G$-magnitude, and $E(B-V)$. The residuals of [$\rm {Fe/H}$] in this work are defined as $\Delta \rm [{Fe/H}]=[\rm {Fe/H}]_{mod}-[\rm {Fe/H}]_{other}$, where $[\rm {Fe/H}]_{other}$ represents the [Fe/H] from the PASTEL or SAGA catalogs. Errors in the predicted metallicities are evaluated by the mean deviation (“bias”) and 1$\sigma$ uncertainties, which are estimated using Gaussian fits. There is no significant bias in the training and the testing sets between the uncertainty-aware CSNet results and the catalogs from PASTEL and SAGA. The uncertainties of $\Delta \rm [{Fe/H}]$ for the reference catalog from PASTEL and SAGA are 0.09 dex and 0.22 dex, respectively. The low levels of bias and uncertainty between the uncertainty-aware CSNet predictions and SAGA values for stars with $\rm [Fe/H]<-1.0$ suggest that the trained model performs well on metal-poor stars.

To estimate the errors, we divided the samples into different [Fe/H] bins. Figure \ref{fig:err_branch} shows comparisons of the median predicted errors and the standard deviations of the residuals for the $G$-magnitude bins from the PASTEL and SAGA catalogs, respectively. Overall, the predicted errors from our model align well with the true values. However, a perfect match is not expected, as the uncertainty estimation branch of the uncertainty-aware CSNet uses an unsupervised learning algorithm.

\subsubsection{Comparison of Metallicity Estimates with the Literature} \label{Comparison of Other Surveys}

\begin{figure*}
  \centering
  \includegraphics[width=0.9\textwidth]{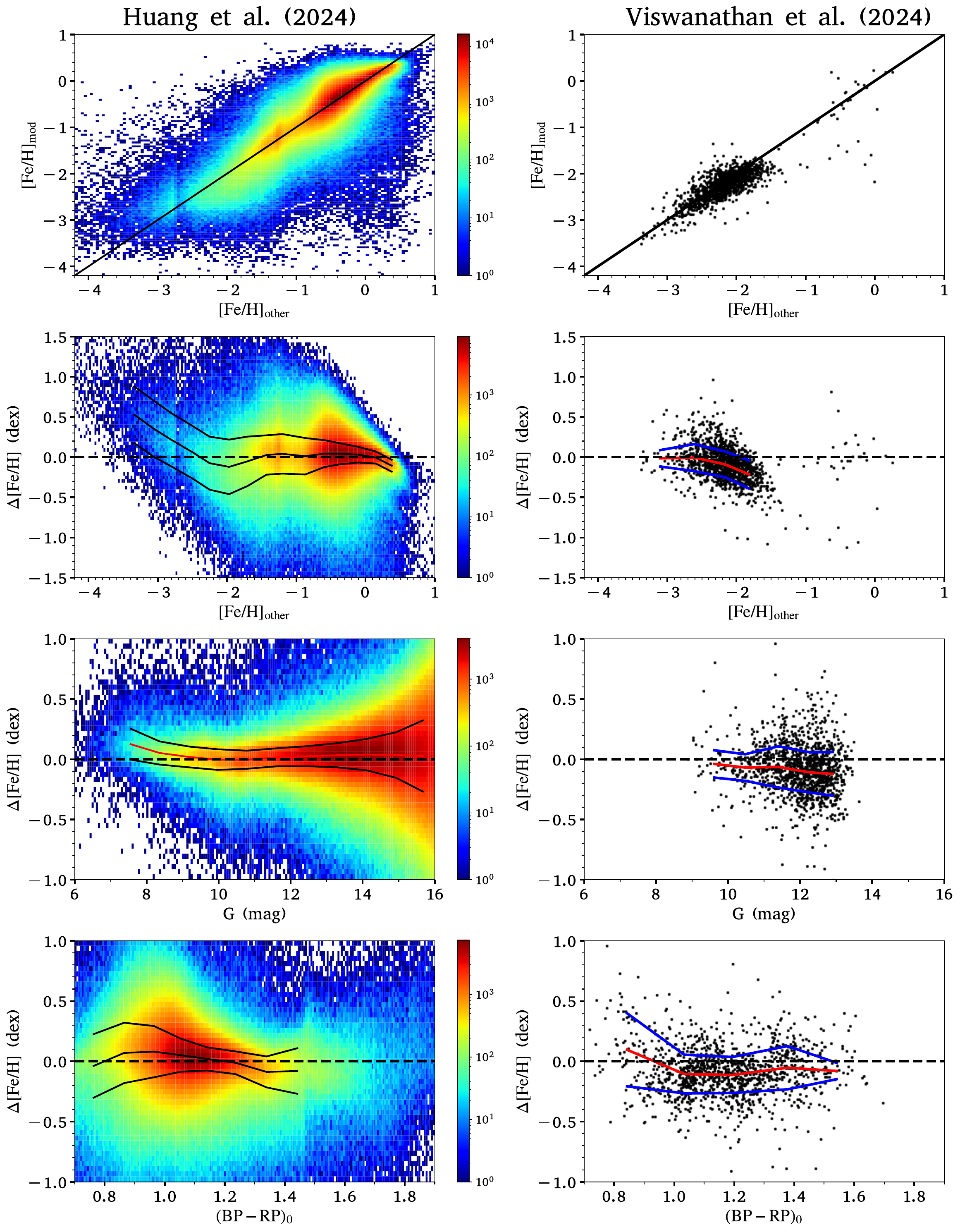}
  \caption{Comparison of [Fe/H] for common stars between this work and \cite{Huang2024} (left panels) as well as \cite{Viswanathan2024} (right panels). The left panels are color-coded by number density. The first row presents our estimates, as a function of [Fe/H], from the two literature sources. The black solid line is the one-to-one line.  The lower three rows display $\Delta$[Fe/H], as a function of [Fe/H], from the two sources, $G$-magnitude, and intrinsic $BP-RP$ color, respectively. The black solid lines represent the mean deviation and the $1\sigma$ uncertainty of the residuals. }
  \label{fig:cmp_GaiaXP}
\end{figure*}

\begin{figure*}
  \centering
  \includegraphics[width=0.9\textwidth]{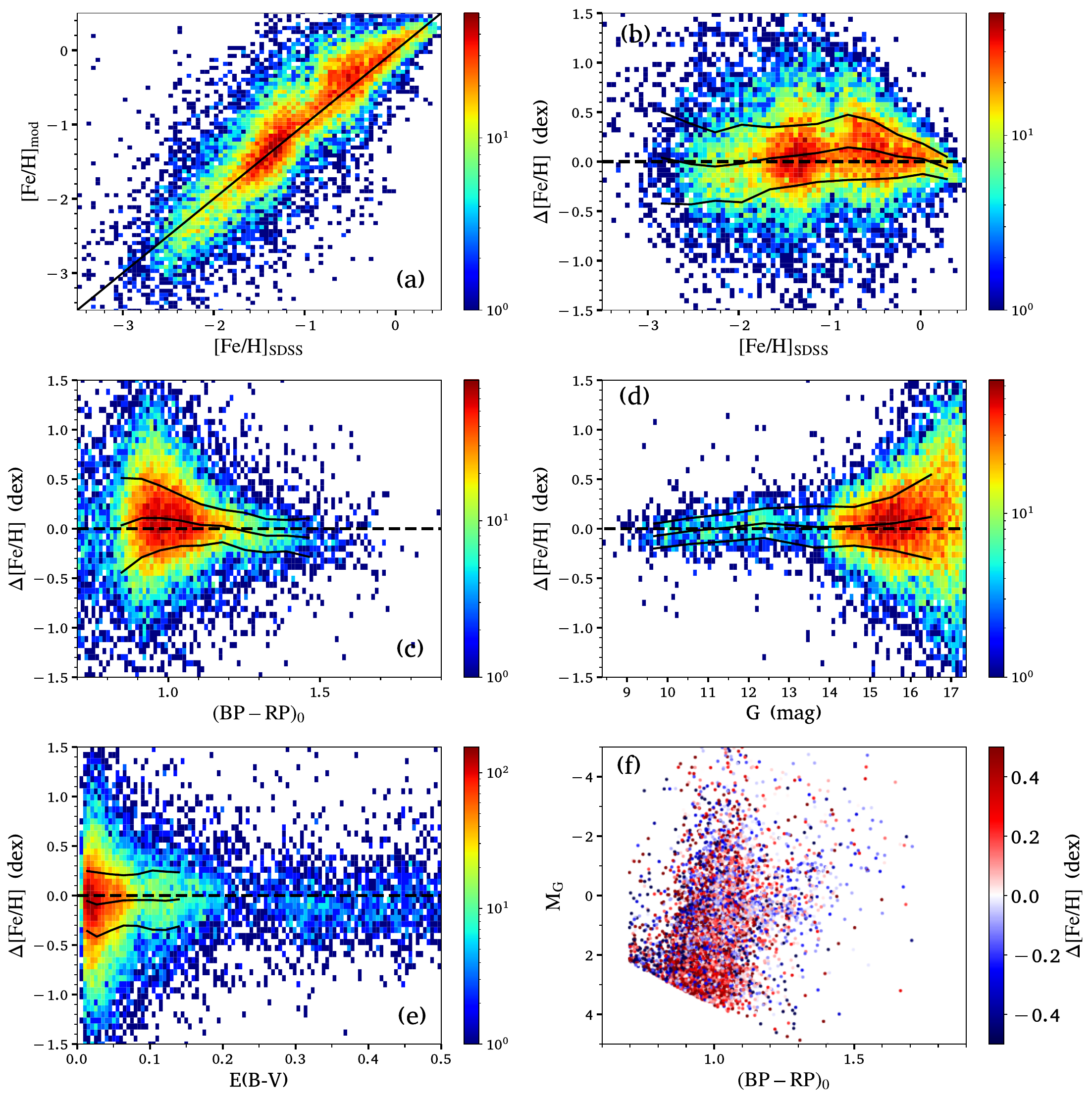}
  \caption{Comparison of [Fe/H] between this work and the SDSS/SEGUE catalog. (a) Our estimates, as a function of [Fe/H], from the SDSS/SEGUE catalog. The black solid line is the one-to-one line. (b)--(e) $\Delta$[Fe/H], as a function of $\rm [Fe/H]_{SDSS}$, intrinsic $BP-RP$ color, $G$-magnitude, and $E(B-V)$, respectively. The black solid lines represent the mean deviation and the $1\sigma$ uncertainty of the residuals. (f) The HR diagram, color-coded by $\Delta$[Fe/H].}
  \label{fig:Yang_X_SDSS}
\end{figure*}

\begin{figure*}
  \centering
  \includegraphics[width=0.9\textwidth]{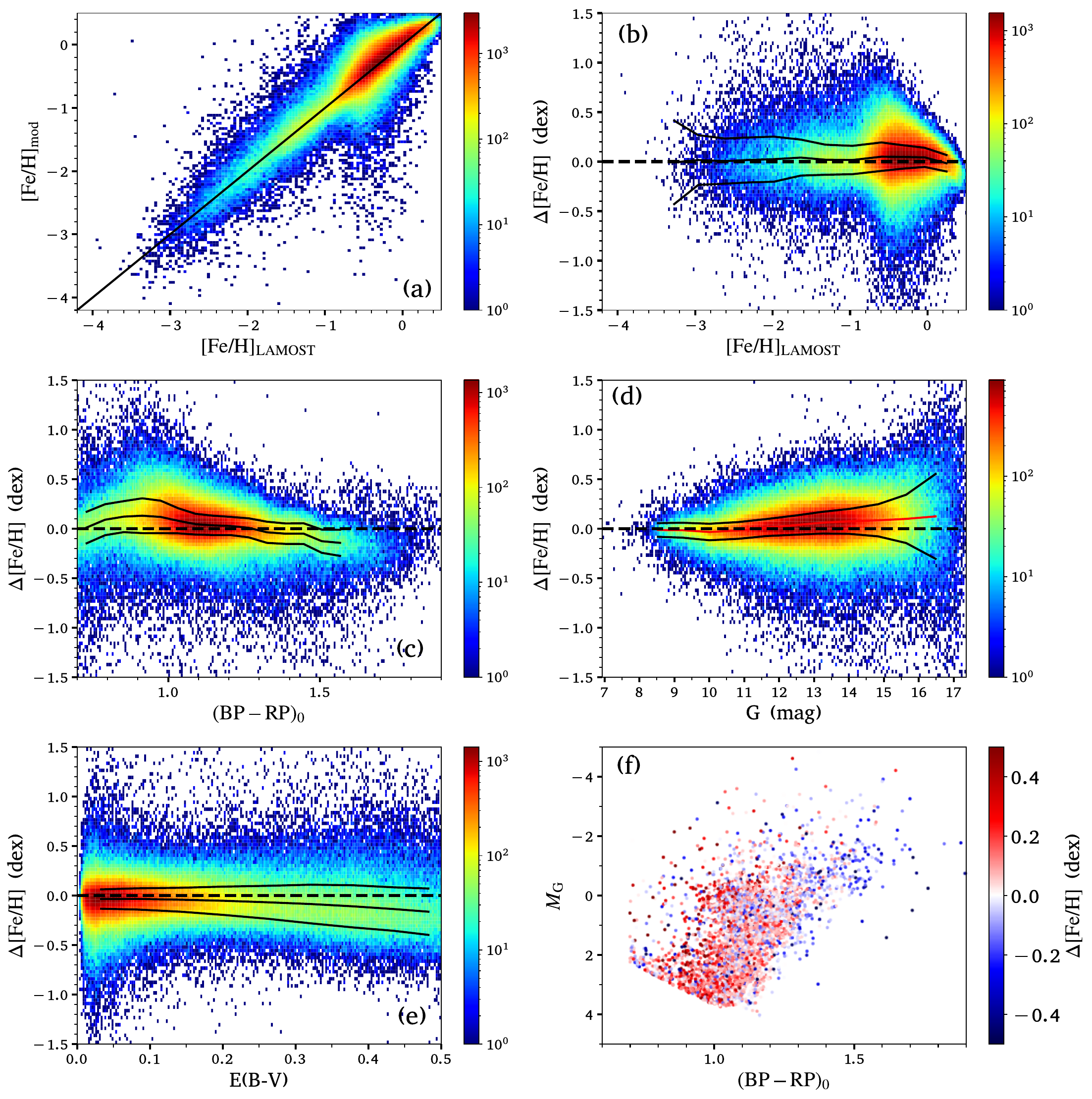}
  \caption{Similar to Figure \ref{fig:Yang_X_SDSS}, but for the comparison of [Fe/H] between this work and LAMOST DD-{\it payne} catalog.}
  \label{fig:Yang_X_LMDR9}
\end{figure*}

To further verify the accuracy and reliability of our model, we compare our results with two Gaia catalogs of \cite{Huang2025} and \cite{Viswanathan2024}, as well as spectroscopic catalogs from 
SDSS/ SEGUE (\citealp{Alam2015}) and LAMOST DD-{\it Payne} (Zhang et al. submitted).
The results will be detailed individually. To ensure the reliability of the comparison results, only sources that meet the first three criteria of the training set, and have $E(B-V)<0.5$ are included.

\cite{Huang2025} provided photometric metallicities for 100 million stars based on synthetic Gaia colors. Following their recommendation, we cross-matched our catalog with theirs for stars with $FLAG = 11111$ and $Region>112$. To ensure the quality of our results, we further required stars with $G < 16$, resulting in a final common sample of 3,024,458 stars. \cite{Viswanathan2024} provided a metallicity catalog for over a thousand bright ($G<13$) VMP stars from Gaia. We found 1276 stars in common. Figure \ref{fig:cmp_GaiaXP} shows the comparisons between our results and the two reference catalogs mentioned above. As shown in the left panel of Figure \ref{fig:cmp_GaiaXP}, the [Fe/H] estimates from \cite{Huang2025} are consistent with ours across different $G$-magnitudes and intrinsic $BP-RP$ colors, except for stars with [Fe/H]$<-2.75$, where a noticeable systematic trend is observed. 
This discrepancy can be attributed to two possible reasons.
One is the non-Gaussian uncertainty associated with their photometric metallicity estimation in the regime of extremely low sensitivity.
The other is underestimation in their results, arising from their assumption that metallicity sensitivity progressively diminishes at the metal-poor end and ultimately drops to zero at [Fe/H] = $-$5.25 (see their Section 3.3). While these assumptions provide robust discrimination in the extreme low-sensitivity regime, they may introduce artificial systematic biases that require correction. 
The density discontinuity of their results at [Fe/H] = $-$2.75 arises from the limited range of the correction applied to the Hertzsprung-Rusell Diagram (HRD; see their Section 4.2). This correction is more effective at reducing contamination for stars with [Fe/H] $> -$2.75 than for those with [Fe/H] $< -$2.75. A minor discontinuity of 0.1 dex at $BP-RP\sim 1.45$ is clearly observed in the lower left panel of Figure \ref{fig:cmp_GaiaXP}, which is from the catalog of \cite{Huang2025}. In that work, two different polynomial models were applied for stars with $BP-RP \leq 1.45$ and $BP-RP>1.45$, respectively. Given that the stellar loci exhibit minimal sensitivity to [Fe/H] around $BP-RP\sim 1.45$, \cite{Huang2025} incorporated $M_G$ for stars with $BP-RP>1.45$ when deriving [Fe/H].
The right panel of Figure \ref{fig:cmp_GaiaXP} shows that our results are in close agreement with those of \cite{Viswanathan2024}, with a typical scatter of 0.17 dex for VMP stars. The better consistency with \cite{Viswanathan2024} is attributed to the improved precision as the sources become brighter.

The SDSS DR12 marks the final data release of SDSS-III, encompassing all observations up to July 2014. Its parameter catalog contains around 400,000 sources with three fundamental stellar parameters: effective temperature ($T_{\rm eff}$), surface gravity ($\log g$), and metallicity ([Fe/H]). These parameters were derived from the Sloan Extension for Galactic Understanding and Exploration (SEGUE) Stellar Parameter Pipeline (\citealp[SSPP;][]{Allende2008,Lee2008a,Lee2008b,Lee2011,Lee2013}), covering a temperature range of 4000 to 10,000 K. We cross-matched our catalog with the SDSS/SEGUE catalog and found 18,478 stars in common. 

The LAMOST DR9 release includes over 10 million low-resolution ($R \sim 1800$) stellar spectra. Zhang M. et al. (submitted) estimated stellar parameters and elemental abundances for 8 million stars from these spectra using the DD-{\it Payne} method (\citealp{Xiang2019}). We cross-matched  our catalog with their LAMOST DR9 catalog, and obtained 433,128 stars in common. 

Figures \ref{fig:Yang_X_SDSS} and \ref{fig:Yang_X_LMDR9} show the comparisons of [Fe/H] values between our estimates and those from the two reference catalogs mentioned above.  Our results demonstrate good agreement with those provided by the SDSS/SEGUE and LAMOST catalogs down to [Fe/H] $\sim -3.5$ . At [Fe/H] $<-2.0$, the typical scatter between our results and the SDSS/SEGUE and LAMOST catalogs are 0.35 and 0.25 dex, respectively. In Figures \ref{fig:Yang_X_SDSS}(c) and \ref{fig:Yang_X_LMDR9}(c), a weak systematic trend across the $BP-RP$ color is observed. We examined the differences between the PASTEL catalog and the two reference catalogs and found a consistent systematic trend, suggesting that this trend likely inherits the systematic discrepancies between the training set and LAMOST/SDSS catalogs. Note that the metallicity uncertainty is higher for bluer stars because their spectra exhibit fewer metallic features, as shown in Figure \ref{fig:sample_space}. As shown in panels (b)--(e), the $\Delta \rm [Fe/H]$ values exhibit no discernible trends with $\rm [Fe/H]_{LAMOST}$, $G$-magnitude, or $E(B-V)$. 
Panels (f)  display $\Delta$[Fe/H] in the HR diagram.
To avoid crowding, only 30\% and 10\% of the common sources are plotted, respectively.
Overall, $\Delta$[Fe/H] has no significant structure in the $\rm (BP-RP)_0$--$M_G$ plane, except for redder sources ($BP-RP)_0 > 1.5$), due to the inherited systematic trend with dereddened $BP-RP$ color.

\subsubsection{Tests with Star Clusters} 
\begin{figure*}
  \centering
  \includegraphics[width=\textwidth]{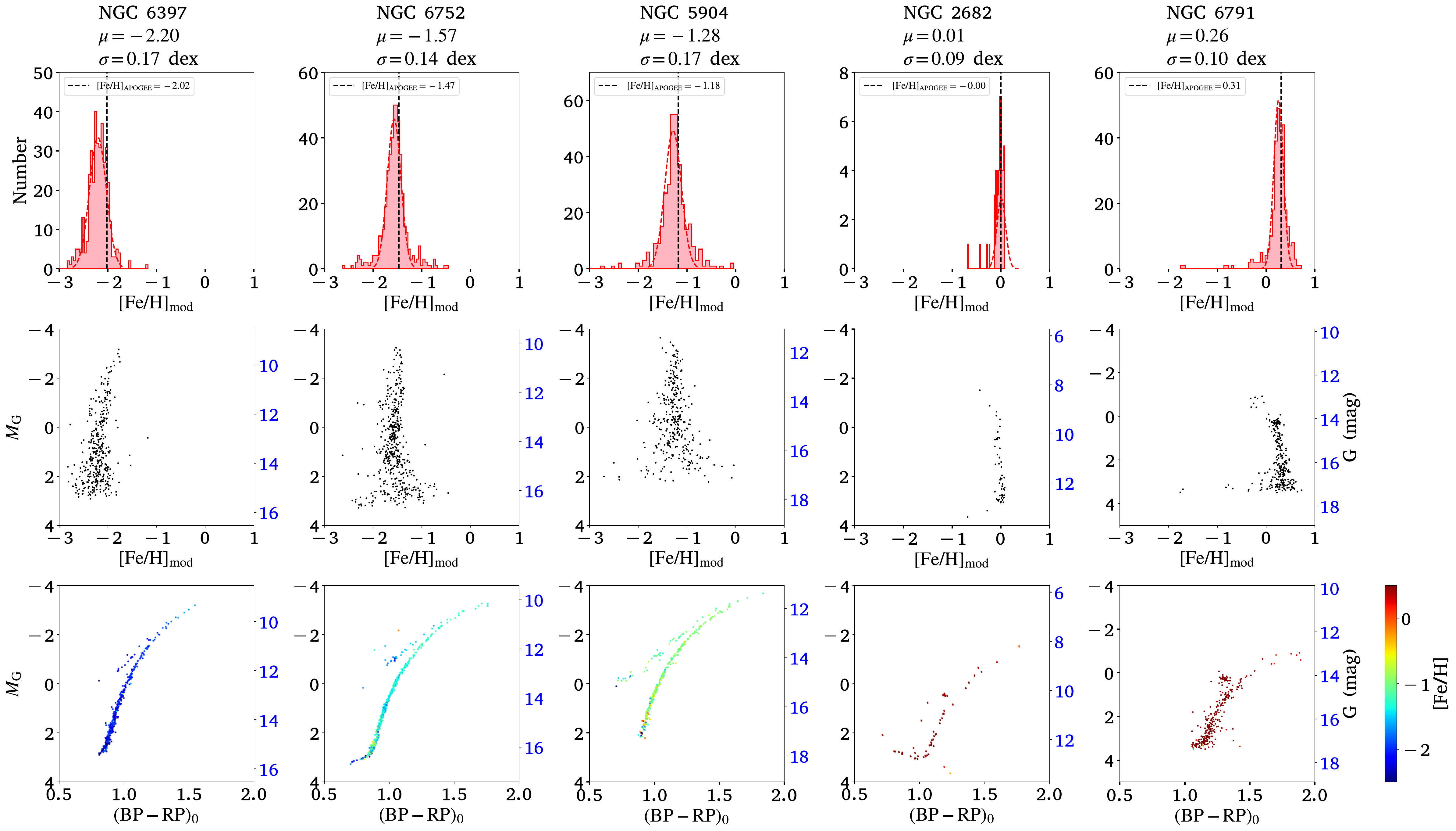}
  \caption{Tests with five clusters. The title of each column lists the cluster name, along with the mean and $1 \sigma$ of the predicted [Fe/H] values, estimated using a Gaussian fit with $3-\sigma$ clipping. The first row shows the distribution of metallicity estimates for member stars of each cluster from our results. The $\rm [Fe/H]_{APOGEE}$ values are taken from APOGEE DR17 (\citealp{Majewski2017}). The second row displays the [Fe/H] distribution from our results, varying with $M_G$ and $G$, represented by the left and right vertical axes, respectively. 
  The third row of panels shows the HRDs of these clusters, with the colors of the dots denoting $\rm [Fe/H]_{\rm mod}$, as coded on the color bar on the right. The vertical coordinates are the same as those in the second row.}
  \label{fig:cluster_test}
\end{figure*}
In this section, we test our model on member stars from three globular clusters (NGC 6397, NGC 6752, and NGC 5904) and two open clusters (NGC 2682 and NGC 6791). The three globular clusters are selected from the catalog compiled by \cite{Baumgardt2021}, while the two open clusters are taken from \cite{Jaehnig2021}. We cross-matched these member stars with our catalog, applying the first three criteria for the selection of the training set in Section \ref{training and tesing sets}. To obtain a pure sample of giant stars, we further require that the member stars of NGC 6791 satisfy $(BP-RP)_0 > 1.05$.

The results are presented in Figure \ref{fig:cluster_test}. For NGC 6397, the mean value of $\rm [Fe/H]_{mod}=-2.20$, with a dispersion of 0.17 dex, reasonably
matches the literature value of $\rm [Fe/H]_{APOGEE}=-2.02$. For NGC 6752, the mean value of $\rm [Fe/H]_{mod}=-1.57$, with a dispersion of 0.14 dex, also in good agreement with literature value of $\rm [Fe/H]_{APOGEE}=-1.47$. For NGC 5904, the mean value of $\rm [Fe/H]_{mod}=-1.28$, with a dispersion of 0.17 dex, is consistent with the literature value of $\rm [Fe/H]_{APOGEE}=-1.18$. For NGC 2682, a dispersion of 0.09 dex, centered around $\rm [Fe/H]_{mod}=+0.01$, is obtained for our model. The mean value closely matches $\rm [Fe/H]_{APOGEE}=0.00$, as reported in APOGEE DR17. For NGC 6791, the mean value of $\rm [Fe/H]_{mod}=+0.26$, with a dispersion of 0.10 dex, is also consistent with the literature value of $\rm [Fe/H]_{APOGEE}=+0.31$. Note that the slight difference between our [Fe/H] values and those from the APOGEE catalog is primarily inherited from the discrepancies between the APOGEE catalog and the training data.

\subsubsection{The Influence of Carbon Enhancement} 

\begin{figure*}
  \centering
  \includegraphics[width=\textwidth]{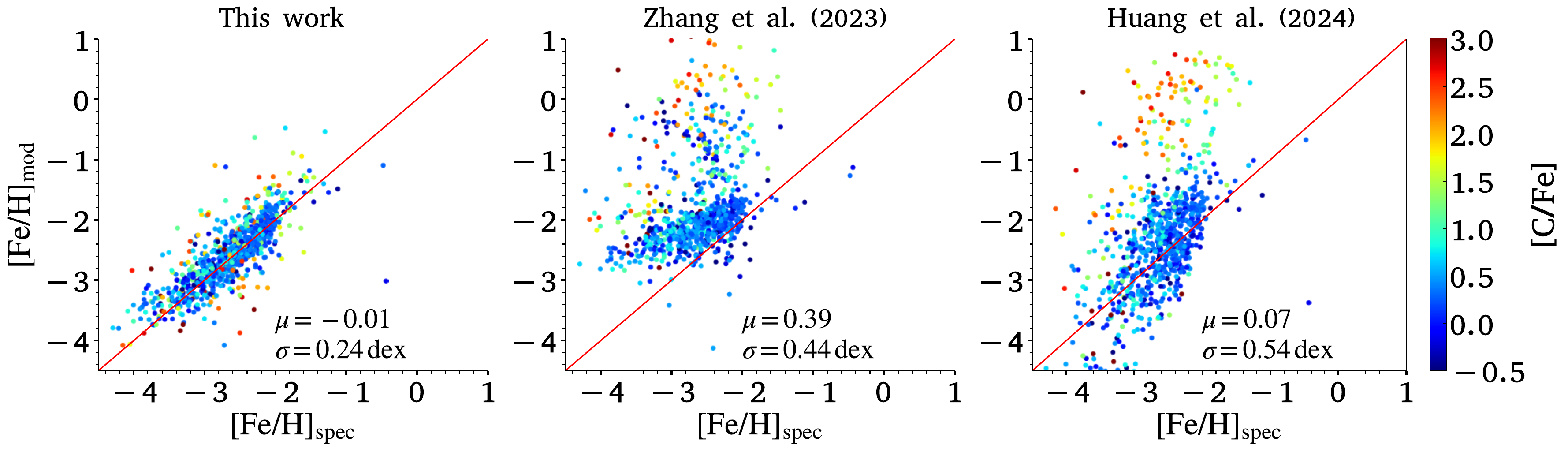}
  \caption{Comparisons of [Fe/H] between the three Gaia catalogs and the one with spectroscopic measurements for VMP/EMP stars (taken from the compilation of \citealt{Hong2024}), color-coded by [C/Fe].  The solid red lines are the one-to-one lines. In the left panel, note the low scatter, symmetrical distribution of points around the one-to-one line, and insensitivity to [C/Fe] for the derived [Fe/H].}
  \label{fig:cfe_influence}
\end{figure*}

Photometric metallicity estimates for VMP or EMP stars are often overestimated due to the contamination of blue narrow/medium-band filters by molecular bands of carbon and nitrogen(\citealp{Huang2023,Hong2024, Lu2024}). In this section, we checked whether our [Fe/H] estimates of are influenced by carbon enhancement. We cross-matched our Gaia DR3 catalog with the medium- and high-resolution spectroscopic catalog of VMP and EMP stars, which includes [Fe/H] and [C/Fe] measurements compiled by \cite{Hong2024}. After applying the first three cuts used for the training set, we obtained 771 unique stars in common. We also performed the same test on the results from \cite{Zhang2023} and \cite{Huang2025}.

Figure \ref{fig:cfe_influence} presents a comparison between our results and those from the other two catalogs, color-coded by [C/Fe] values. The sources shown in the three panels are identical.  The $\rm [Fe/H]_{spec}$ and [C/Fe] values in the catalog of \cite{Hong2024} are originally from \cite{Yoon2016}, \cite{Li2022}, \cite{Placco2022}, and \cite{Zepeda2023}. As shown, our estimates are in good agreement with the $\rm [Fe/H]_{spec}$ values, achieving a precision of $\rm \delta [Fe/H] = 0.24$ dex. Notably, stars with high [C/Fe] values are symmetrically distributed around the one-to-one lines in our results, indicating that our metallicities are not overestimated, in contrast to the overestimation observed in the catalogs from \cite{Zhang2023} and \cite{Huang2025}. We conclude that our results are reliable for VMP and EMP stars, even in cases of significant carbon enhancements. There are two possible reasons for the good [Fe/H] estimate achieved with our method. First, the Gaia SEDs contain information about carbon. Second, our training sample includes a diverse range of stars, both influenced by C-enhancement and not.

\subsection{The Final Sample} \label{Gaia DR3 catalog}

\begin{figure}
  \centering
  \includegraphics[width=0.5\textwidth]{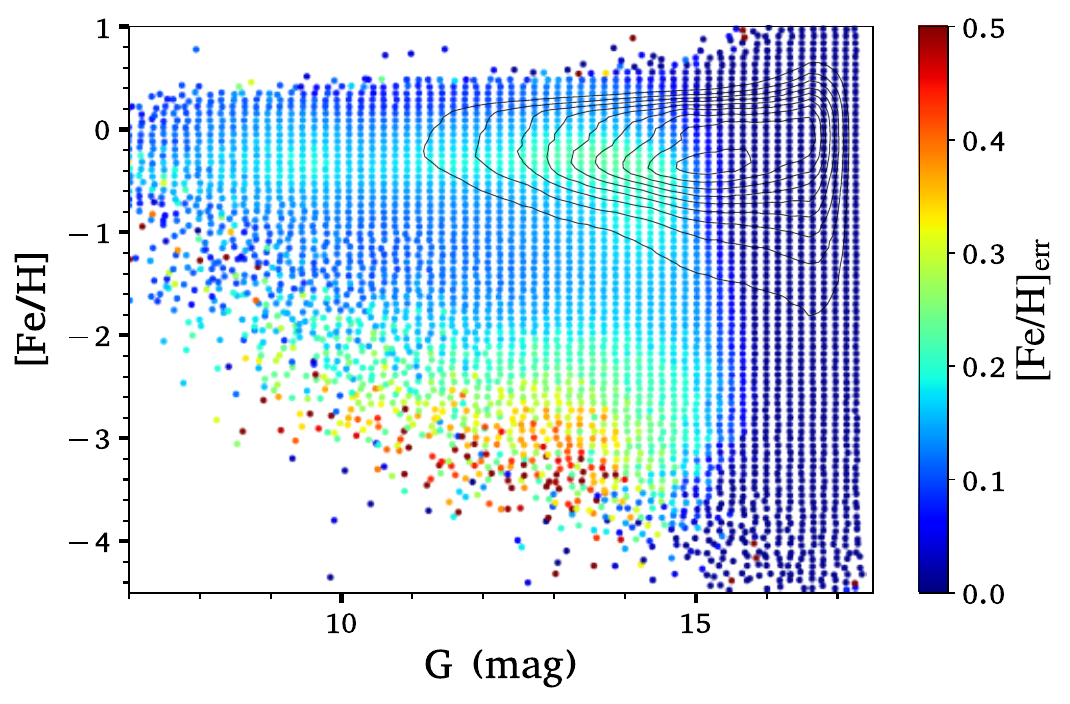}
  \caption{Distribution of [Fe/H] uncertainties in the $G$-- [Fe/H] plane for the final sample.}
  \label{fig:feh_err}
\end{figure}

\begin{table*}
  \centering
  \caption{Description of the Final Sample}
  \begin{threeparttable}
  \label{table:table2}
  \begin{tabular}{lll}
  \hline
   Col. &Field &    Description  \\
  \hline        
   1& source\_id &   Unique source identifier for Gaia XP spectra (unique with a particular data release) \\
   2& gl &   Galactic longitude (deg)\\
   3& gb &  Galactic latitude (deg)  \\ 
   4& ra &   Right ascension (deg)\\
   5& dec &  Declination (deg)  \\
   6& phot\_bp\_rp\_excess\_factor &  BP/RP excess factor from Gaia DR3  \\  
   7& BP\_RP0 &  Intrinsic $BP-RP$ color after color correction   \\
   8& PHOT\_G\_MEAN\_MAG &  $G$-band photometry from Gaia DR3  \\ 
   9& M\_G &  Absolute magnitude in the $G$-band from Gaia DR3   \\  
   10& EBV &  Interstellar extinction from the \citet{Schlegel1998} dust-reddening map \\
   11& rgeo &  Geometric distances from \cite{Bailer2021}   \\
   12& FEH &  Metallicity derived from the the uncertainty-aware CSNet   \\ 
   13& FEH\_err &  Estimated error of [Fe/H] \\
   \hline
  \end{tabular}

  \end{threeparttable}
\end{table*}

\begin{figure*}
  \centering
  \includegraphics[width=0.8\textwidth]{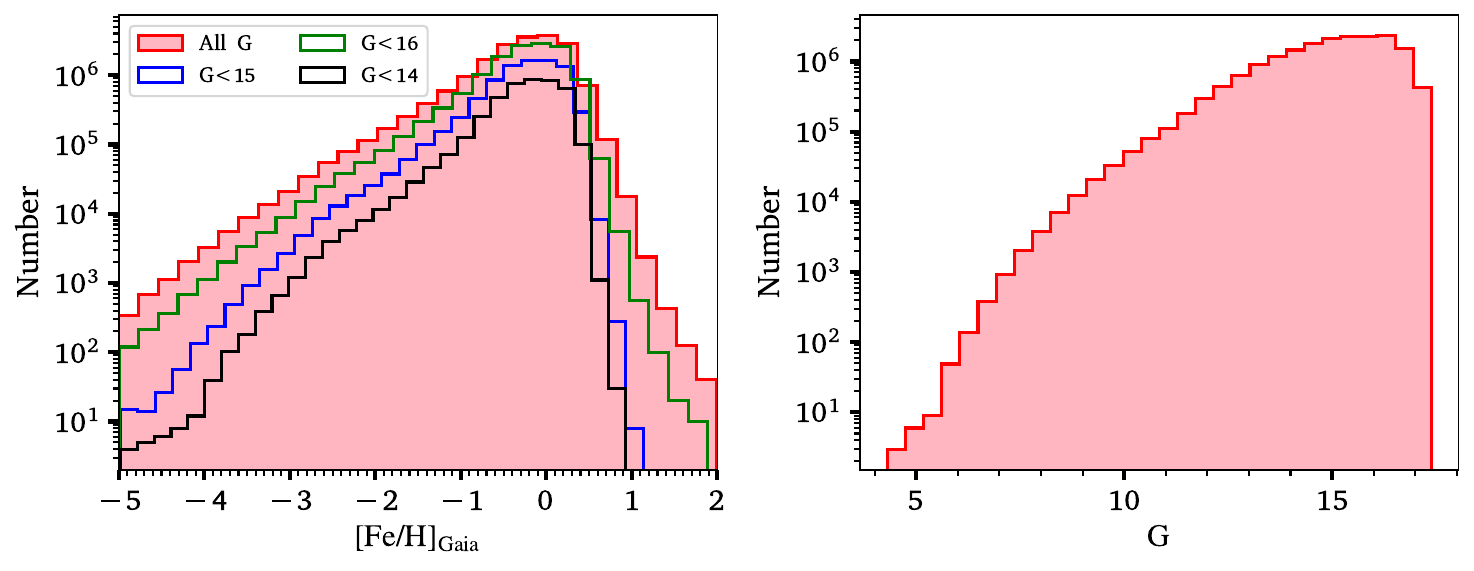}
  \caption{The [Fe/H] and $G$-magnitude distributions for the final sample.}
  \label{fig:final_sample}
\end{figure*}

\begin{figure*}
  \centering
  \includegraphics[width=\textwidth]{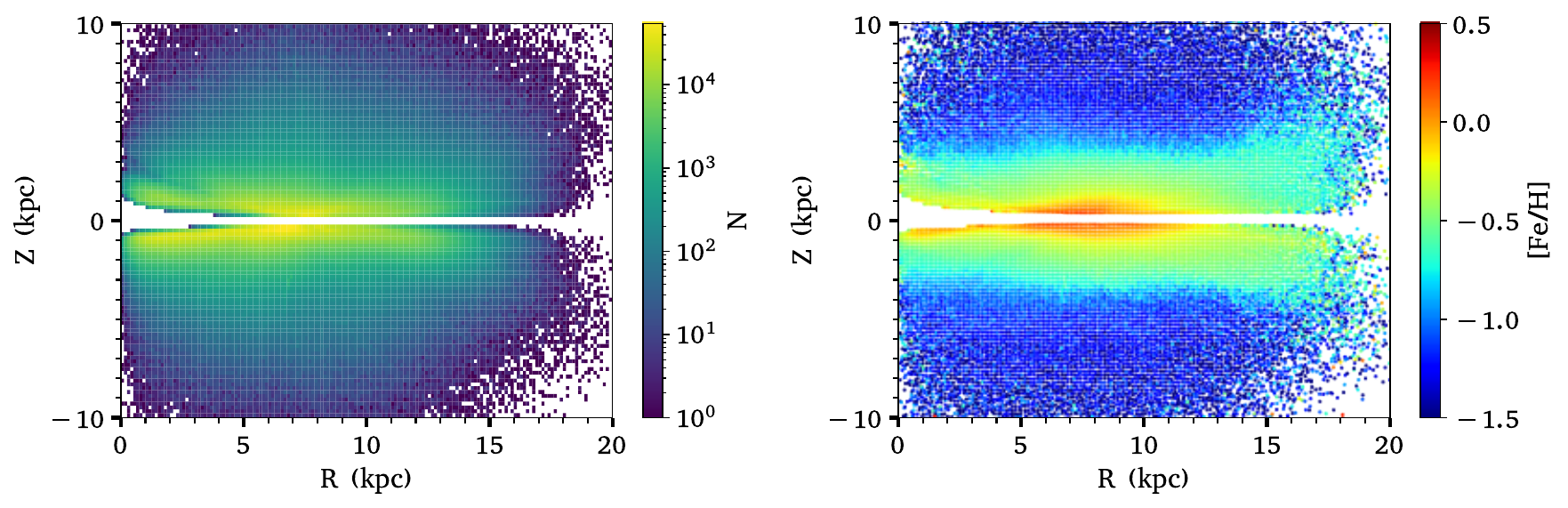}
  \caption{Spatial distributions of the final sample in the Z--R plane.  The left panel is color-coded by number density, while the right panel is color-coded by [Fe/H].}
  \label{fig:r_z}
\end{figure*}

After applying the first three criteria for the selection of the training set in Section \ref{training and tesing sets}, our final sample includes 31,360,788 giant stars from Gaia XP sources. We then applied the trained uncertainty-aware CSNet to this sample.
The uncertainties of our results are presented in Figure\,\ref{fig:feh_err}. Given the limitation of the uncertainty-aware CSNet, we recommend using predicted errors only for stars with $G<15$ to avoid extrapolation, as this is consistent with the range of the training samples. We confirm the reliability of the predicted errors by comparing them with those obtained from star clusters, finding good agreement. Note that the noticeable increase in predicted errors around $\rm [Fe/H] \sim -0.3$ is attributed to the insufficient training samples in this region, leading to higher epistemic uncertainty. The catalog is publicly available, and a description of the final sample catalog is provided in Table\,\ref{table:table2}.

To ensure photometric quality and correct extinction, we further selected 20 million stars with $E(B-V)_{\rm SFD}<0.5$ mag as the final reliable sample. Their [Fe/H] and $G$-magnitude distributions are shown in Figure\,\ref{fig:final_sample}, aligning with the expected trend.
Note the final sample contains 1,089,712 VMP and 369,269 EMP candidate stars.

The spatial distributions of the number density and the determined metallicities in the R--Z plane for the final sample are shown in Figure\,\ref{fig:r_z}, where $Z$ represents the distance to the Galactic plane, and $R$ is the Galactocentric distance. Evident vertical gradients in metallicity among the disk-system stars are observable from the Galactic plane.

\section{Summary} 
\label{summary}

In this work, we present UA-CSNet, a model for estimating metallicities from dereddened and corrected Gaia BP/RP (XP) spectra for giant stars. Using the PASTEL catalog as the training sample, our model is capable of predicting metallicities down to $\rm [Fe/H] \sim -4$. Besides providing estimates of metallicity, the UA-CSNet can also deliver uncertainties associated with the predicted values. These predicted uncertainties are validated using star clusters. Though these uncertainties are only applicable for stars with $G<15$ to avoid extrapolation, the intermediate predicted uncertainties generated during each iteration of the training process continuously improve the model's predictive accuracy for [Fe/H].

We validate our metallicity estimates by comparing them with external catalogs, including SAGA, LAMOST DD-{\it Payne}, SDSS/SEGUE, and two Gaia catalogs, as well as through tests on star clusters, finding overall strong agreement. Unlike metallicity estimates from other studies based on Gaia XP spectra, which tend to be overestimated due to molecular band contamination in blue filters, our estimates for VMP and EMP stars are unaffected by carbon enhancement.
Applying UA-CSNet, we obtain reliable and precise metallicity estimates for approximately 20 million giant stars, including 0.36 million VMP stars and 0.05 million EMP stars. 
The catalog of the estimated metallicities is publicly accessible.

\vspace{7mm} \noindent {\bf Acknowledgments}

This work is supported by the National Key R\&D Program of China via  2024YFA1611601 and 2024YFA1611901, and the National Natural Science Foundation of China through the projects NSFC 12222301, 12173007, and 124B2055. M.X. acknowledges financial support from the National Key R\&D Program of China through grant no. 2022YFF0504200.
T.C.B. acknowledges partial support from Physics Frontier Center/JINA Center for the Evolution of the Elements (JINA-CEE), and OISE-1927130: The International Research Network for Nuclear Astrophysics (IReNA), awarded by the US National Science Foundation (NSF).

Guoshoujing Telescope (the Large Sky Area Multi-Object Fiber Spectroscopic Telescope LAMOST) is a National Major Scientific Project built by the Chinese Academy of Sciences. Funding for the project has been provided by the National Development and Reform Commission. LAMOST is operated and managed by the National Astronomical Observatories, Chinese Academy of Sciences. 
This work has made use of data from the European Space Agency (ESA) mission {\it Gaia} (https://www.cosmos.esa.int/gaia), processed by the {\it Gaia} Data Processing and Analysis
Consortium (DPAC, https://www.cosmos.esa.int/
web/gaia/dpac/ consortium). Funding for the DPAC has been provided by national institutions, in particular the institutions participating in the {\it Gaia} Multilateral Agreement.
Data resources are supported by China National Astronomical Data Center (NADC) and Chinese Virtual Observatory (China-VO). This work is supported by Astronomical Big Data Joint Research Center, co-founded by National Astronomical Observatories, Chinese Academy of Sciences and Alibaba Cloud.

\end{CJK*}
\end{document}